\documentclass[conference]{IEEEtran}

\usepackage{cite}
\usepackage{amsmath,amssymb,amsfonts}
\usepackage{algorithmic}
\usepackage{graphicx}
\usepackage{textcomp}
\usepackage{xcolor}
\usepackage{booktabs}
\usepackage{array}
\usepackage{tabularx}
\usepackage{makecell}
\usepackage{float}
\def\BibTeX{{\rm B\kern-.05em{\sc i\kern-.025em b}\kern-.08em
    T\kern-.1667em\lower.7ex\hbox{E}\kern-.125emX}}
\begin{document}

\title{Knowledge-Graph-Guided Retrieval-Augmented LLMs for Explainable Root Cause Analysis in Automotive HiL Validation}

\author{\IEEEauthorblockN{1\textsuperscript{st} Hamza Ouarrad*}
\IEEEauthorblockA{\textit{Technische Universität Clausthal} \\
Clausthal-Zellerfeld, Germany \\
hamza.ouarrad@tu-clausthal.de}
*Corresponding author
~\\
\and
\IEEEauthorblockN{2\textsuperscript{nd} Mohammad Abboush}
\IEEEauthorblockA{\textit{Technische Universität Clausthal} \\
Clausthal-Zellerfeld, Germany \\
mohammad.abboush@tu-clausthal.de}

~\\
\and
\IEEEauthorblockN{3\textsuperscript{rd} Andreas Rausch}
\IEEEauthorblockA{\textit{Technische Universität Clausthal} \\
Clausthal-Zellerfeld, Germany \\
andreas.rausch@tu-clausthal.de}
~\\
}

\maketitle

\begin{abstract}
Hardware-in-the-Loop validation of automotive software systems generates large multivariate time-series recordings whose manual analysis is time-consuming and often limited to anomaly detection and fault classification rather than root-cause analysis. Although deep learning methods have shown strong performance in fault detection and classification, they usually require task-specific training or retraining when new fault locations, systems, or operating conditions are introduced. They also tend to treat localization as a classification task, without explicitly representing the spatial and functional relationships between fault locations, sensors, and downstream subsystem effects. This limits their generalizability and their usefulness for engineering root cause analysis and diagnosis. This paper proposes a knowledge-graph-guided retrieval-augmented large language model framework for RCA (root cause analysis) and fault localization in automotive HiL data. The method converts raw time-series recordings into compact diagnostic evidence, enriches this evidence with sensor-to-location and propagation knowledge, and retrieves similar historical cases to support the final reasoning step. The LLM is then used as a decision and explanation layer rather than as a direct time-series classifier, producing a ranked fault-location prediction together with an interpretable RCA explanation. The framework is evaluated on two automotive HiL case studies: an ASM gasoline engine and an electric vehicle system. The best-performing model achieves Top-1 accuracies of 90\% and 94\%, respectively, while recording-level aggregation reaches perfect file-level fault localization in the evaluated subset. These results demonstrate the potential of KG-guided RAG-LLM reasoning for explainable and generalizable HiL RCA.
\end{abstract}

\begin{IEEEkeywords}
Hardware-in-the-Loop, fault localization, root-cause analysis, large language models, retrieval-augmented generation, knowledge graph, automotive software validation, time-series diagnosis
\end{IEEEkeywords}

\section{Introduction}
In the automotive domain, the development of safety-critical software systems is governed by the ISO 26262 functional safety standard \cite{url}, which prescribes a rigorous verification and validation process across the V-model life cycle, collectively referred to as X-in-the-Loop (XiL) testing \cite{abboush2022intelligent}. In the concluding validation phase preceding series production, real test drives on public roads, complemented by hardware-in-the-loop (HIL) simulations integrating actual electronic control units (ECUs) and physical vehicle components, are conducted to validate the implemented system against its functional safety requirements under representative operating conditions \cite{abboush2024virtual,abboush2025explainable}.
A direct consequence of modern HIL validation campaigns is the generation of large-volume test recordings that manifest as multivariate time series data with nonlinear dynamics and multi-mode behaviour. Industry-certified tools predominantly rely on rule-based analysis and predefined thresholds, which are effective for flagging anomalies but provide little support for identifying the actual location and root cause of a fault. In practice, an anomaly observed in a single signal may originate from sensors, communication channels, software components, controllers, or actuators \cite{salih2021fault}, and the corresponding fault effects often propagate through interconnected components before becoming visible in the recordings \cite{wen2003case}. Consequently, validation engineers must manually inspect large numbers of signals and combine numerical observations with prior experience, historical fault cases, component knowledge, sensor dependencies, and system architecture information. There is therefore a pressing need for intelligent diagnostic frameworks that integrate data-driven evidence with engineering knowledge to support explainable and efficient root cause analysis in HIL validation.

Within the literature, signal-based \cite{olsson2004fault}, knowledge-based \cite{venkatasubramanian2003review}, and model-based \cite{isermann2005model} approaches have been proposed for the analysis of test recordings, yet each suffers from well-known limitations related to noise sensitivity, manual engineering effort, and dependence on reference model fidelity. More recently, machine learning (ML) and deep learning (DL) methods have demonstrated remarkable performance in fault detection and diagnosis (FDD) by automatically learning discriminative representations from data \cite{liu2025intelligent,lin2025fd}. Nevertheless, irrespective of the underlying paradigm, existing approaches generally answer only whether and what type of anomaly has occurred, and do not address the more critical question faced by validation engineers: where the fault is located and why a given location constitutes the most probable root cause \cite{wen2003case}.

Recent advances in Large Language Models (LLMs) have opened new opportunities for knowledge-intensive reasoning in engineering diagnosis \cite{zhang2025large,qi2025large}, while Retrieval-Augmented Generation (RAG) techniques \cite{lewis2020retrieval} have been successfully employed for industrial troubleshooting and knowledge retrieval \cite{ma2025knowledge,zhang2025_large}. However, most existing LLM-based diagnostic approaches treat the language model as an end-to-end classifier and do not incorporate structured diagnostic evidence or fault localization knowledge into the reasoning process \cite{lukens2025evaluation}. Furthermore, the literature lacks knowledge-guided retrieval-based frameworks tailored to fault localization in automotive HIL environments: existing methods rarely integrate numerical time-series evidence, historical fault cases, sensor-to-component mappings, and propagation relationships within a unified reasoning framework \cite{xu2025iterative}, and most retrieval-based LLM systems rely on semantic embeddings rather than engineered diagnostic evidence derived from multivariate time-series behaviour \cite{lewis2020retrieval}. In addition, LLM-based diagnostic systems remain susceptible to hallucinations and model-specific biases, which limits their direct deployment in safety-critical contexts without additional grounding mechanisms \cite{xu2025iterative,lin2025automotive}. Consequently, a significant research gap exists at the intersection of automotive HIL validation, retrieval-augmented reasoning, case-based diagnosis, structured knowledge integration, and LLM-assisted fault localization.

To address this gap, the present article proposes a novel knowledge-guided retrieval-augmented LLM framework for root cause analysis and fault localization in the real-time HIL validation of automotive software systems. The framework transforms raw multivariate HIL recordings into structured diagnostic evidence, including abnormality scores, signal reaction times, deviation directions, and inter-signal correlation changes. These engineered features drive a lightweight case-based retrieval mechanism, in contrast to conventional embedding-based retrieval \cite{lewis2020retrieval,lukens2025evaluation}, and are combined with structured domain knowledge describing sensor roles, component relationships, and fault propagation paths \cite{ma2025knowledge}. Unlike conventional LLM-based classification approaches, the LLM acts as an evidence-fusion, reasoning, and reranking layer rather than as a direct classifier: it receives the structured evidence, retrieved cases, domain knowledge, and candidate fault locations, and produces a ranked list of the most probable root causes accompanied by interpretable engineering explanations, supporting both single and combined fault hypotheses. To the best of the authors' knowledge, this is the first study that unifies automotive HIL recordings, structured diagnostic evidence, case-based retrieval, knowledge-guided diagnosis, root cause analysis, and Retrieval-Augmented Large Language Models within a single framework.
The principal contributions of this work are summarised as follows:
\begin{itemize}
\item	A knowledge-guided retrieval-augmented LLM framework is proposed for fault localization and root cause analysis in automotive HIL validation. 
\item	A diagnostic evidence extraction module converts raw HIL recordings into interpretable features, including abnormality scores, reaction times, deviation directions, and correlation changes. 
\item	A lightweight case-based retrieval approach combined with domain knowledge captures sensor-location relationships and fault propagation patterns. 
\item	An LLM-based reasoning layer integrates evidence, retrieved cases, and engineering knowledge to generate ranked fault-location hypotheses with root-cause explanations.

\end{itemize}
The remainder of this article is organised as follows. Section 2 reviews the related literature. Section 3 presents the proposed approach. Section 4 describes the implementation structure and case study. Section 5 reports and discusses the experimental results. Section 6 concludes the article and outlines directions for future research.

\section{Related work}

\subsection{Deep Learning-Based Fault Detection and Diagnosis in Automotive Systems}

Data-driven FDD in the automotive domain has progressed from shallow classifiers anchored in hand-engineered features towards deep architectures that learn discriminative temporal representations directly from raw multivariate signals. Convolutional--dense pipelines have been applied to sensor-fault detection, isolation and prognostic health-index forecasting in autonomous-driving stacks, with Safavi et al. \cite{safavi2021multi} reporting 99.84\% detection accuracy on the Audi A2D2 dataset but restricting evaluation to four canonical fault classes injected via a static normal-distribution model. Recurrent architectures have dominated diagnostics in electrified powertrains, exemplified by the LSTM model of Kaplan et al. \cite{kaplan2021fault}, which reduced MAPE from 10.13\% to 2.06\% over a shallow ANN baseline, although fault injection was performed at the Simulink level rather than under real-time constraints. Ensemble and hybrid formulations have broadened coverage and robustness: a two-class/one-class ensemble attaining an F2-score of 77\% on OBD-II road-trial recordings under known and unseen faults \cite{theissler2017detecting}; a multi-label LSTM--Random Forest ensemble augmented by a GRU-based denoising autoencoder reaching 99.43\% accuracy and 91.2\% F1 on concurrent sensor faults under HIL-generated noise \cite{abboush2023intelligent}; and a Random Forest classifier coupled with sliding-mode fault-tolerant control on a TruckMaker-based HIL platform for heavy-duty brakes \cite{raveendran2020brake}; complementary reviews further note that many algorithms still simplify actuator dynamics and neglect parameter drift, limiting transferability to production conditions \cite{pietrowski2024fault}. Collectively, these contributions advance detection and classification accuracy but answer only whether and what type of anomaly has occurred, providing no explicit support for identifying where a fault is located or why a particular location is the most probable root cause — a gap that has motivated two parallel responses in the literature: knowledge-augmented language models and dedicated root-cause-diagnosis methods.

\subsection{Knowledge Graph Enhanced Large Language Models for Fault Diagnosis}

The integration of Knowledge Graphs (KGs) with Large Language Models (LLMs) has emerged as a leading paradigm for grounding LLM reasoning in structured engineering knowledge and mitigating hallucinations in industrial fault diagnosis. A first line treats the KG as an external memory queried by an LLM agent: To-FD-EKG \cite{men2025interpretable} couples a table-sequence triplet extractor with an LLM that interacts with a virtual digital-twin environment through perception--action loops, reaching 82.3\% average cosine similarity on wind-turbine maintenance logs, whilst LC-CoT \cite{li2026long} extends the paradigm to compound faults by combining causal hypergraphs with reinforcement-learning fine-tuning. A second line embeds graph structure directly into the LLM through prefix-tuning and subgraph generation, illustrated by the joint KG--LLM aviation-assembly system of \cite{peifeng2024joint}, which reaches 98.5\% accuracy on industrial localisation cases, and by joint-extraction frameworks raising entity-relation F1 from 0.907 to 0.968 \cite{li2025knowledge}.

A third and increasingly dominant line employs Retrieval-Augmented Generation over a domain KG. CNC-focused systems consolidate diagnostic cases, work orders and equipment data into KGs queried by fine-tuned LLMs, outperforming experienced engineers on benchmark tasks \cite{nie2026industrial}; wind-turbine pipelines combine DeepSeek-R1 with dual-channel graph and vector retrieval and Graph Attention Networks for multi-hop reasoning \cite{wang2025wt}; LLM-FDQA reports 96.1\% Hit@1 entity linking and 94.04\% end-to-end diagnostic accuracy on real-world CNC data through explicit evidence-chain reasoning \cite{gao2026knowledge}; Graph-RAG for train-bogie diagnosis achieves 92.67\% attribution consistency through hybrid sparse--dense--graph retrieval \cite{miao2025graph}; and dynamic multimodal frameworks fuse vibration spectrograms with text KGs through DeepSeek-V3, sustaining 85.3\% accuracy at $-5$ dB SNR\cite{zhuang2025large}. Causal extensions for automotive chip production further couple knowledge graphs with text-grounded causal event graphs to support long-chain reasoning \cite{liu2026llm}. Despite this momentum, three limitations recur. Validation is overwhelmingly concentrated in process industries---wind turbines, CNC machinery, aviation and rail--- with virtually no exposure to automotive HIL recordings. Retrieval remains predominantly embedding-based and operates on unstructured maintenance logs, leaving the use of engineered diagnostic features derived from multivariate time-series behaviour largely unexplored. And the LLM is most often deployed as an end-to-end classifier or generator rather than as an evidence-fusion and reranking layer grounded in sensor-to-component relationships and fault-propagation paths.

\subsection{Root Cause Diagnosis of Industrial Faults}

Root cause diagnosis (RCD) of industrial faults has been pursued along three complementary directions. The first relies on causal discovery over multivariate process recordings: CASCN \cite{zheng2025root} combines LASSO-based fault isolation with a channel-attentive sparse causal TCN, achieving on the Tennessee Eastman process (TEP) a 23.10\% improvement in true-positive rate and a 60.82\% reduction in false-positive rate over mainstream baselines; REASON \cite{wang2023interdependent} integrates hierarchical graph neural networks for topological causal discovery with Extreme Value Theory for individual-entity scoring; MPRA combines Lasso--Granger single-level causality with attention-based high-level similarity for grouped RCD \cite{zhao2025causal}; and extended convergent cross mapping with redundancy pruning \cite{ji2025explainable}, reinforcement-learning structure learning \cite{pan2026temporal}, and improved symbolic transfer entropy \cite{wang2018information} extend causal discovery to aero-engine, chemical and electromechanical processes.

A second direction couples anomaly detection with post-hoc attribution to recover causes from effects. Deep Root Cause Analysis \cite{huang2024deep} pairs hierarchical prediction networks with saliency maps and explicitly demonstrates on TEP that high-deviation variables often correspond to symptoms rather than direct causes; LSTM-AVAGMM with Kernel SHAP supports anomaly explanation and incremental adaptation in wind-turbine SCADA streams \cite{zhang2024research}; VAE--RCAP--SHAP separates internal-sensor anomaly manifestation from external operational drivers on container-ship engines \cite{park2026explainable}; physics-guided graphs fuse explicit knowledge rules with multivariate GNNs for backtracking root causes across wind farms, coal pulverisers and sugar factories \cite{feng2024root}; and a data-space attention-CNN with LSTM--OLS captures causal propagation paths in float-glass and TEP \cite{qiao2024root}. A third, knowledge-centric direction integrates KGs and case-based reasoning with anomaly detection, exemplified by an informed self-supervised one-class learner coupled with FMEA-grounded SPARQL, symbolic-driven neural reasoning and case-based retrieval for smart-factory predictive maintenance \cite{klein2025combining}. Within the automotive domain, evidence is limited: a causal Seq2Seq with attention \cite{huang2025analysis} addresses fault identification and basic causal relations on UCI/Ford GoBike/Nissan datasets, but does not localise faults to physical components under HIL constraints.

Across these contributions, three challenges remain unresolved. Causal-discovery methods deliver structural transparency but depend on offline training over substantial fault corpora and offer limited grounding in engineering knowledge such as sensor-to-component mappings and propagation paths. Attribution-based methods supply local explanations yet frequently confuse symptoms with causes and rarely support multi-fault hypotheses. Knowledge-graph and case-based approaches incorporate expert structure but have not been integrated with retrieval-augmented LLM reasoning over engineered time-series evidence, and automotive HIL validation under safety-critical concurrent-fault conditions remains absent altogether. These observations motivate the present investigation, which unifies structured diagnostic evidence extracted from HIL recordings, lightweight case-based retrieval over engineered features, structured sensor--component--propagation knowledge, and an LLM reasoning-and-reranking layer into a single framework for fault localisation and root cause analysis in automotive HIL validation.

\section{Methodology}

\begin{figure*}[t]
    \centering
    \includegraphics[width=\textwidth]{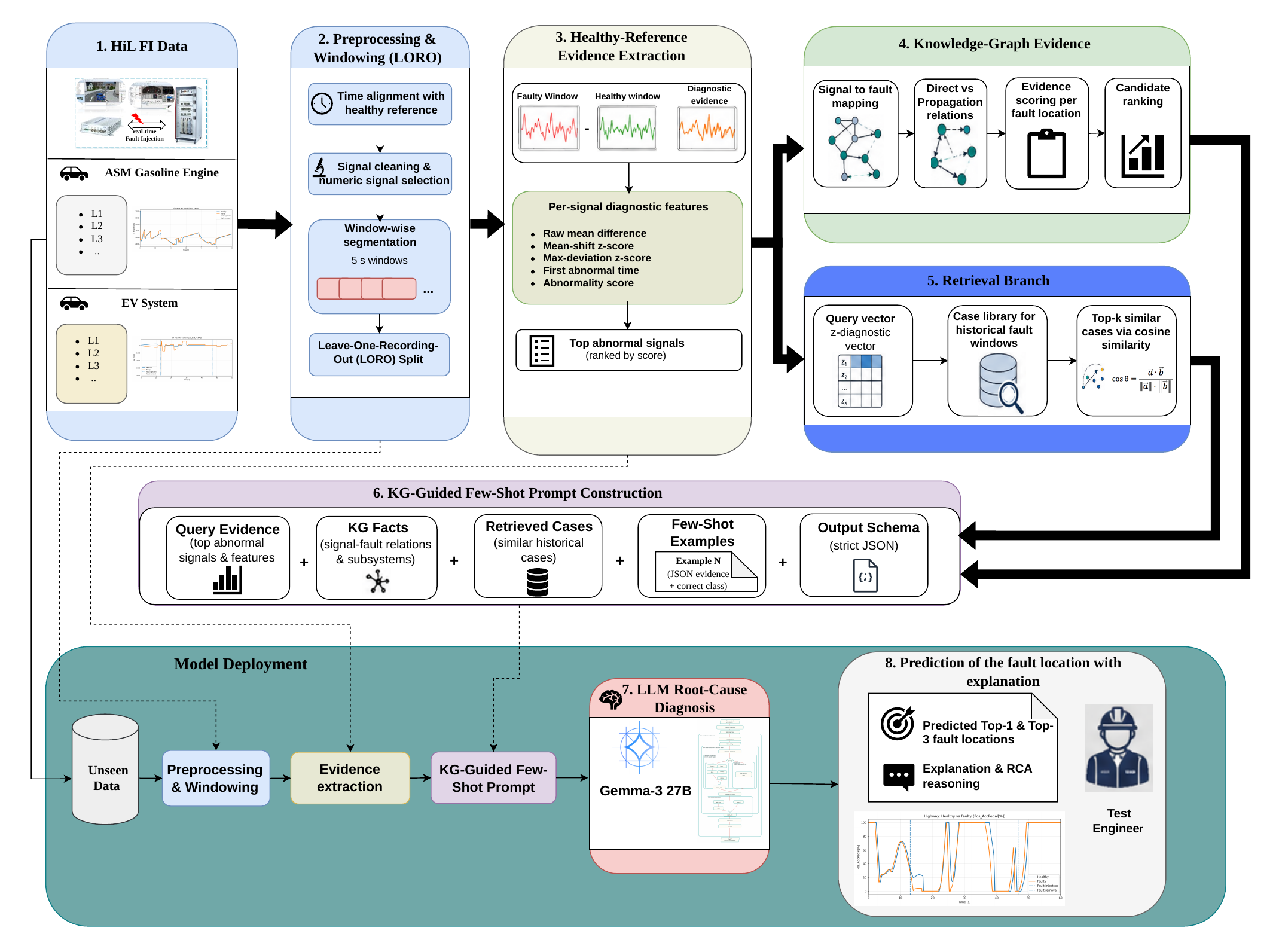}
    \caption{Overview of the proposed KG-guided RAG RCA workflow for HiL time-series data.}
    \label{fig:rag_kg_llm_methodology}
\end{figure*}

The proposed method is designed for root-cause diagnosis in multivariate temporal HiL time-series data. Given a recording $X \in \mathbb{R}^{T \times d}$, where $T$ denotes the number of time steps and $d$ the number of sensor signals, the objective is to identify the most likely fault location from a predefined set of candidates. Fig.~\ref{fig:rag_kg_llm_methodology} summarizes the complete KG-guided RAG pipeline. The method first converts raw recordings into compact diagnostic evidence, then combines knowledge-graph-based candidate reasoning with retrieval of similar historical cases, and finally uses an LLM to produce a ranked fault-location prediction with an explanatory root-cause diagnosis.

\subsection{Preprocessing and Healthy-Reference Evidence Extraction}
\label{subsec:preprocessing_evidence}

The raw HiL recordings are first converted into fault-relevant diagnostic evidence. Each experimental run is loaded as a multivariate CSV time series and sorted according to the time column. Non-signal fields, such as the time stamp and optional status labels, are excluded from the diagnostic feature set. Since the measurements are generated under closed-loop driving conditions, absolute signal values alone are not sufficient for fault localization. The same signal may vary naturally because of the driving profile, controller response, or vehicle dynamics. Therefore, each faulty recording is evaluated relative to a corresponding healthy reference run.

The analysis focuses on the fault-active region of the experiment. A pre-fault interval is used as baseline behavior, while the stable fault interval is used to describe the abnormal response. In this work, the baseline window is defined as 2--11~s and the stable fault interval as 15--45~s. For window-wise diagnosis, the stable fault interval is further segmented into fixed-length temporal windows. This preserves the temporal structure of the fault response and allows the method to capture whether a signal reacts early, remains persistently abnormal, or appears mainly as a later propagation effect.

For each signal \(x_j\), the faulty recording is compared with the time-aligned healthy reference over the same interval. Let \(x_j(t)\) denote the query signal and \(x_j^{h}(t)\) the corresponding healthy reference signal. The raw mean deviation for a window \(W\) is computed as
\begin{equation}
\Delta \mu_j(W)=
\frac{1}{|W|}\sum_{t \in W} x_j(t)
-
\frac{1}{|W|}\sum_{t \in W} x_j^{h}(t)
\label{eq:raw_mean_deviation}
\end{equation}
This quantity captures the direction and magnitude of the deviation in the original signal units. To make signals with different scales comparable, the deviation is also normalized using a robust reference standard deviation. A small standard-deviation floor is used to avoid unstable z-scores for nearly constant HiL signals:
\begin{equation}
z_j(W)=
\frac{|\Delta \mu_j(W)|}
{\max(\sigma_j^{h}(W), \epsilon)}
\label{eq:mean_shift_z}
\end{equation}
In addition to the mean-shift score, the maximum deviation from the healthy reference is computed in normalized form. The first abnormal reaction time is also estimated by detecting the first time point at which the reference deviation exceeds a threshold. This timing feature is important for root-cause reasoning: a direct fault-location signal that reacts early is usually more diagnostic than a downstream signal that becomes abnormal only after propagation. The extracted evidence for each window therefore contains raw mean difference, mean-shift z-score, maximum-deviation z-score, first abnormal reaction time, and an overall abnormality score. The abnormality score combines persistent deviation and peak deviation, while clipping extreme values to prevent a single transient spike from dominating the ranking. The resulting evidence table is sorted by abnormality score to obtain the most relevant direct and propagated signal deviations. This compact representation is then passed to the knowledge-graph and retrieval stages instead of sending the full raw time series to the LLM.

\subsection{KG-Guided Candidate Ranking and RAG}
\label{subsec:kg_candidate_ranking_rag}

After the diagnostic evidence has been extracted, the next step is to translate abnormal signal behavior into plausible fault-location candidates. This is done using a knowledge graph that links each candidate fault location to its physically related signals. The graph contains three types of information: signal-to-fault mappings, direct and propagation relationships, and diagnostic rules. Direct mappings define which signal represents a possible injected fault location, such as accelerator-pedal position for an accelerator-pedal fault or battery voltage for a high-voltage battery fault. Propagation relations describe how a fault at one location may affect downstream signals through the vehicle dynamics, powertrain, or electrical subsystem.

For each candidate fault location \(c\), a KG-guided evidence score is computed from primary, direct, and propagation signal evidence:
\begin{equation}
S_{\mathrm{evid}}(c)
=
w_p S_p(c)
+
w_d S_d(c)
+
w_g S_g(c)
\label{eq:kg_evidence_score}
\end{equation}
where \(S_p(c)\), \(S_d(c)\), and \(S_g(c)\) denote the primary-signal, direct-signal, and propagation-signal evidence, respectively. The weights give the strongest importance to primary and direct KG evidence, while propagation signals provide supporting evidence only. This design reflects the diagnostic assumption that downstream effects can confirm a propagation path, but they should not dominate the root-cause decision when the direct fault-location signal provides clear evidence.

Retrieval in inference time is performed by comparing the query evidence vector with reference evidence vectors from the historical case library. The similarity between a query vector \(\mathbf{v}_q\) and a reference vector \(\mathbf{v}_r\) is computed using cosine similarity:
\begin{equation}
\mathrm{sim}(\mathbf{v}_q,\mathbf{v}_r)
=
\frac{\mathbf{v}_q^{\top}\mathbf{v}_r}
{\|\mathbf{v}_q\|_2 \|\mathbf{v}_r\|_2}
\label{eq:cosine_similarity}
\end{equation}
where \(\mathbf{v}_q\) is the query evidence vector and \(\mathbf{v}_r\) is a reference vector from the historical case library.

To avoid retrieval leakage, a leave-one-recording-out strategy is applied. When a query recording is evaluated, all windows from the same recording are removed from the retrieval library and from the few-shot example pool. The retrieved cases therefore come only from other recordings, although they may contain the same fault location under a different gain or severity. This setting is important because the method should generalize to unseen recordings rather than retrieve near-duplicate windows from the same experiment.

The final candidate ranking combines KG-guided evidence with retrieval support. For each candidate \(c\), the score is computed as
\begin{equation}
S(c)
=
\alpha S_{\mathrm{evid}}(c)
+
(1-\alpha)S_{\mathrm{ret}}(c)
\label{eq:final_candidate_score}
\end{equation}
where \(S_{\mathrm{ret}}(c)\) is obtained from the retrieved cases and \(\alpha\) controls the relative influence of direct diagnostic evidence and retrieval. In the ASM case study, retrieval is used as a supporting signal beside strong KG-based evidence, while in the EV case study its influence is kept smaller because the available EV reference library is more limited. The ranked candidates are then passed to the prompt-construction stage together with the extracted evidence, KG facts, retrieved reference cases, and the structural relations between fault locations, directly affected sensors, and downstream subsystems. These spatial and functional relationships are important because faults in HiL systems do not only appear as isolated signal deviations; they can propagate through physically connected components, such as from driver inputs to engine behavior or from battery-side deviations to DC-link and propulsion responses. In this way, the retrieval component does not simply retrieve similar examples. It augments the LLM input with structured historical evidence that is constrained by the knowledge graph, the subsystem relationships, and the candidate-ranking mechanism.

\subsection{Few-Shot Prompt Construction, LLM Reasoning, and Prediction Aggregation}
\label{subsec:prompt_llm_aggregation}

At inference time, each query recording or query window is processed through the same evidence-extraction and candidate-ranking pipeline as the historical cases. The query evidence vector is compared with the historical case library, and the top-(k) most similar cases are retrieved. To avoid label leakage, the query filename and hidden ground-truth label are excluded from the prompt, since file names contain fault-location information. The LLM therefore receives only observable diagnostic evidence, KG facts, retrieved reference cases, few-shot examples, and the ranked candidate locations.

The prompt is constructed as a compact diagnostic report rather than as raw time-series input. It includes the query identifier, analyzed time interval, strongest direct-location signal deviations, relevant propagation evidence, KG/domain facts, retrieved similar cases, few-shot decision examples, and the calibrated candidate ranking. The few-shot examples are selected from historical reference windows and exclude windows from the same recording as the query. Their purpose is to demonstrate the expected reasoning pattern: primary and direct KG evidence should guide the root-cause decision, while propagation effects and retrieved cases provide supporting context. The model is instructed to return valid JSON only. The output contains the predicted Top-1 fault location, the Top-3 ranked candidate list, a confidence level, candidate-level reasoning, the most likely propagation path, uncertainty notes, and a short RCA explanation. This structure makes the output automatically parsable while still useful for test engineers, who receive not only a predicted label but also an explanation of the supporting evidence and possible downstream effects.

After inference, the response is parsed and validated by checking the JSON structure, required fields, and predicted labels against the predefined fault-location set. For window-wise experiments, the LLM produces one prediction per evaluated fault window. These predictions are aggregated into a recording-level diagnosis using rank-based voting over the Top-3 lists, where higher-ranked candidates receive stronger support. The final file-level prediction is the candidate with the highest aggregated support, which reduces the influence of locally ambiguous windows and provides a more stable diagnosis for the complete HiL recording.

\section{Case Study and Data Description}

\subsection{ASM Gasoline Engine}
\label{subsec:asm_gasoline_engine}

The first case study is based on the ASM gasoline engine system from dSPACE, integrated into a closed-loop automotive Hardware-in-the-Loop (HiL) environment. The setup combines a detailed gasoline-engine plant model with vehicle dynamics, driver/environment models, and a controller layer, enabling fault-injection experiments under realistic real-time operating conditions \cite{abboush2022hardware}. The plant is executed on the dSPACE SCALEXIO real-time platform, while the control logic is deployed through a dSPACE MicroAutoBox II, which emulates the ECU behavior during the experiments.

The gasoline-engine model is considered as part of a coupled vehicle system rather than as an isolated engine component. It includes several interacting subsystems, such as the air path, fuel system, piston engine, exhaust path, and cooling system. This coupling is important for the proposed root-cause localization task because a fault injected at one location can propagate to physically related signals in other subsystems. In this study, five fault locations are considered for the ASM gasoline-engine case: accelerator pedal, brake pedal, steering wheel, engine speed, and throttle position. These locations are represented by the signals \textit{Pos\_AccPedal[\%]}, \textit{Pos\_BrakePedal[\%]}, \textit{Angle\_SteeringWheel[deg]}, \textit{n\_Engine[rpm]}, and \textit{Pos\_Throttle[\%]}, respectively.

\subsection{Electric Vehicle System}

\label{subsec:electric_vehicle_system}

The second case study uses an electric vehicle system integrated into the same HiL-based validation concept. This case study complements the gasoline-engine setup by considering a different propulsion architecture, where the relevant fault effects are related to electrical power flow, battery behavior, electric-machine dynamics, and vehicle-level responses. The plant and environment models are executed on the dSPACE SCALEXIO real-time platform, while the controller logic is executed through dSPACE MicroAutoBox II in order to emulate ECU behavior in the closed-loop experiment. The EV model includes the main subsystems required for closed-loop electric powertrain validation, including the high-voltage battery, DC-link, rear electric machine, steering system, vehicle dynamics, and driver/environment model. These subsystems are physically and functionally coupled, so a fault injected in one component may also affect downstream electrical or vehicle-level signals. In this study, three EV fault locations are considered: the high-voltage battery, the rear electric machine speed, and the steering system. These locations are represented by the signals \textit{V\_Bat\_HV[V]}, \textit{omega\_EM\_Rear[rad/s]}, and \textit{Trq\_SteeringWheel[Nm]}, respectively.

\subsection{Dataset Description}
\label{subsec}

The dataset used in this study consists of multivariate time-series recordings generated from automated fault-injection experiments in the ASM gasoline-engine and electric-vehicle HiL setups. Each experimental run is stored as a separate CSV file. The ASM gasoline-engine files contain 15 columns, including the time column and sensor measurements from the driver, vehicle, and engine/powertrain subsystems. The electric-vehicle files contain 17 columns, also including the time column, and cover sensor measurements from the battery, DC-link, propulsion, steering, and vehicle-level subsystems. Each recording contains approximately 7{,}700 samples with a fixed sampling interval of 0.01~s.

The fault type considered in the evaluated dataset is a gain fault. For the ASM gasoline-engine case study, two faulty recordings with different gain values are available for each of the five fault locations, resulting in ten ASM fault-injection runs. For the electric-vehicle case study, one faulty recording is available for each of the three fault locations, resulting in three EV fault-injection runs. In total, the evaluated fault-injection subset therefore contains 13 CSV recordings. The recordings include healthy and faulty operating phases, allowing the injected behavior to be analyzed relative to a time-aligned healthy reference trajectory. The experiments capture the temporal system response before, during, and after fault activation. In the considered recordings, the fault is active approximately between 12~s and 47~s. This structure makes the data suitable for evaluating both window-level fault localization and recording-level aggregation. The ASM gasoline-engine case study considers five fault locations: accelerator pedal, brake pedal, steering wheel, engine speed, and throttle position. The electric-vehicle case study considers three fault locations: high-voltage battery, rear electric machine speed, and steering system. The recorded signals therefore provide system-level observations that can be used to evaluate fault localization under different propulsion architectures and subsystem interactions.

\section{Results and Discussion}

\subsection{Fault-Localization Performance and Recording-Level Aggregation}


\begin{table}[h]
\caption{Performance comparison of LLMs for KG-guided fault localization on the ASM gasoline engine and EV case studies.}
\label{tab:llm_fault_localization_results}
\begin{center}
\tiny
\renewcommand{\arraystretch}{1.10}
\setlength{\tabcolsep}{1.2pt}

\resizebox{\columnwidth}{!}{%
\begin{tabular}{@{}llcccccc@{\hspace{1pt}}c@{}}
\toprule
\textbf{Scenario} 
& \textbf{Model} 
& \textbf{Top-1 Acc.} 
& \textbf{Top-3 Acc.} 
& \textbf{MRR} 
& \textbf{F1} 
& \textbf{Precision} 
& \textbf{Recall} 
& \textbf{MCC} \\
\midrule

ASM Gasoline Engine 
& Llama-3.3 70B FP8 
& 0.800 & 0.800 & 0.800 & 0.800 & 0.850 & 0.800 & 0.750 \\

& Mistral-Small 3.2 24B 
& 0.850 & 0.950 & 0.883 & 0.839 & 0.870 & 0.850 & 0.820 \\

& DeepSeek-R1-Qwen 32B
& 0.850 & 0.950 & 0.900 & 0.844 & 0.880 & 0.850 & 0.820 \\

& Qwen3 32B
& 0.900 & 1.000 & 0.933 & 0.899 & 0.910 & 0.900 & 0.878 \\

& \textbf{Gemma-3 27B}
& \textbf{0.900} & \textbf{1.000} & \textbf{0.933} 
& \textbf{0.899} & \textbf{0.910} & \textbf{0.900} & \textbf{0.878} \\

\midrule

EV 
& Llama-3.3 70B FP8 
& 0.889 & 1.000 & 0.926 & 0.886 & 0.917 & 0.889 & 0.849 \\

& Mistral-Small 3.2 24B
& 0.833 & 1.000 & 0.888 & 0.823 & 0.887 & 0.833 & 0.783 \\

& DeepSeek-R1-Qwen 32B
& 0.833 & 1.000 & 0.888 & 0.823 & 0.887 & 0.833 & 0.783 \\

& Qwen3 32B
& 0.944 & 1.000 & 0.963 & 0.944 & 0.952 & 0.944 & 0.921 \\

& \textbf{Gemma-3 27B}
& \textbf{0.944} & \textbf{1.000} & \textbf{0.963} 
& \textbf{0.944} & \textbf{0.952} & \textbf{0.944} & \textbf{0.921} \\

\bottomrule
\end{tabular}%
}

\vspace{1mm}
\parbox{\columnwidth}{\scriptsize MRR: mean reciprocal rank; MCC: Matthews correlation coefficient.}
\end{center}
\end{table}

Table~\ref{tab:llm_fault_localization_results} reports the fault-localization performance of the evaluated LLMs on the ASM gasoline engine and EV case studies. The results show that the proposed KG-guided RAG pipeline provides reliable fault-location ranking across both systems, despite the different signal structures and fault mechanisms. In the ASM gasoline engine case, Gemma-3 27B and Qwen3 32B achieve the strongest performance, with a Top-1 accuracy of 0.900, a Top-3 accuracy of 1.000, and an MRR of 0.933. This means that the correct injected fault location is always contained in the three highest-ranked candidates and is predicted as the first candidate in most windows. In the EV case study, the same two models again obtain the best results, reaching a Top-1 accuracy of 0.944. These results indicate that the proposed prompting strategy does not depend on a single vehicle architecture, but can transfer to a second HiL system when the corresponding KG mapping and diagnostic signal summaries are available. The high Top-3 accuracy is particularly relevant for root-cause analysis because engineers often need a ranked list of plausible fault locations rather than a single label. The results indicate that the evidence-oriented prompt helps preserve the correct fault location among the leading candidates, even when the Top-1 prediction is affected by propagation effects or ambiguity between physically coupled components.

Differences between models are still visible. Llama-3.3 70B FP8 achieves lower accuracy on the ASM gasoline engine case, with a Top-1 accuracy of 0.800 and a Top-3 accuracy of 0.800. This indicates that, although the model is computationally efficient, it is less reliable in preserving the correct candidate under the KG-guided prompt. Mistral-Small 3.2 24B and DeepSeek-R1-Qwen 32B show intermediate performance, especially on the ASM case, where both reach a Top-1 accuracy of 0.850 and a Top-3 accuracy of 0.950. In contrast, Gemma-3 27B and Qwen3 32B consistently achieve the strongest ranking behavior across both case studies. Since Gemma-3 27B reaches the same accuracy level as Qwen3 32B while requiring lower average inference time, it is treated as the most favorable model for the proposed pipeline in the subsequent analysis.

\begin{table}[h]
\caption{Leave-one-recording-out window-level and file-level fault localization results. File-level predictions are obtained by aggregating window-level predictions over each recording.}
\label{tab:loro_window_file_aggregation}
\begin{center}
\tiny
\renewcommand{\arraystretch}{1.12}
\setlength{\tabcolsep}{1.8pt}

\resizebox{\columnwidth}{!}{%
\begin{tabular}{@{}llrrccccc@{}}
\toprule
\textbf{Case Study} 
& \textbf{Model} 
& \textbf{Win.} 
& \textbf{Files} 
& \textbf{Win. T1} 
& \textbf{Win. T3} 
& \textbf{Win. MRR} 
& \textbf{File T1} 
& \textbf{File MRR} \\
\midrule

ASM Gasoline Engine 
& Gemma-3 27B 
& 60 
& 10 
& 0.833 
& 0.917 
& 0.869 
& \textbf{1.000} 
& \textbf{1.000} \\

\midrule

EV 
& Gemma-3 27B 
& 18 
& 3 
& 0.944 
& 1.000 
& 0.963 
& \textbf{1.000} 
& \textbf{1.000} \\

\bottomrule
\end{tabular}%
}

\vspace{1mm}
\parbox{\columnwidth}{\scriptsize Win.: windows; T1: Top-1 accuracy; T3: Top-3 accuracy; MRR: mean reciprocal rank.}
\end{center}
\end{table}

Table~\ref{tab:loro_window_file_aggregation} further evaluates the effect of aggregating window-level predictions at the recording level. The evaluation follows a leave-one-recording-out strategy: when a query recording is evaluated, all windows from that same recording are excluded from the retrieval library and from the prompt examples. This prevents the model from retrieving or seeing windows from the test recording itself and therefore provides a more realistic estimate of generalization to unseen fault recordings. This step is important because individual windows do not always contain the complete fault signature. Some windows capture the early onset of the fault, while others contain stronger propagated effects after the fault has influenced downstream signals. As a result, a single window can be locally ambiguous. For example, a fault injected into an actuator or engine-related signal may later affect vehicle velocity, throttle behavior, pressure signals, or other coupled measurements. The same effect appears in the EV case, where battery-related faults can propagate to electric-machine or DC-link signals. Therefore, evaluating each window independently is useful for temporal localization, but it can underestimate the reliability of the final diagnosis at the recording level.

The aggregation results show that this ambiguity is reduced when predictions are combined over all windows of the same recording. In the ASM gasoline engine case, the window-level Top-1 accuracy is 0.833, while the file-level Top-1 accuracy reaches 1.000. Similarly, in the EV case, the window-level Top-1 accuracy is 0.944, and the file-level Top-1 accuracy also reaches 1.000. This shows that the final decision becomes more stable when multiple fault-active windows contribute to the prediction. The result is relevant for HiL-based fault localization because a recorded experiment is usually analyzed as a complete test run rather than as an isolated window. Consequently, the recording-level aggregation step aligns the evaluation more closely with the intended diagnostic use case: identifying the most likely injected fault location for a full HiL experiment.

\subsection{Retrieval Representation and KG-Guided Evidence}

\begin{figure}[!h]
\centering
\includegraphics[width=0.92\columnwidth]{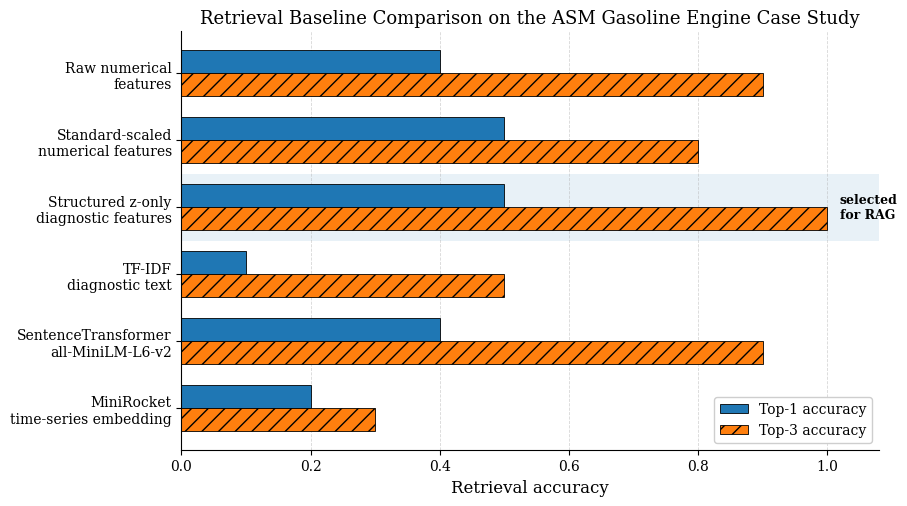}
\caption{Retrieval baseline comparison on the ASM gasoline engine case study. The structured z-only diagnostic representation achieves the strongest Top-3 retrieval accuracy and is selected for the final KG-guided RAG pipeline.}
\label{fig:retrieval_baseline_comparison}
\end{figure}

The retrieval component was evaluated independently to identify the most suitable representation for retrieving diagnostically similar HiL fault windows. As shown in Fig.~\ref{fig:retrieval_baseline_comparison}, the structured z-only diagnostic representation achieves the strongest performance on the ASM gasoline engine case study, with 0.50 Top-1 accuracy and 1.00 Top-3 accuracy. This representation uses healthy-reference mean-shift and maximum-deviation z-scores, which makes deviations comparable across signals with different physical units and dynamic ranges. In contrast, raw numerical vectors, text-based representations, sentence embeddings, and MiniRocket embeddings show weaker retrieval performance. The final KG-guided RAG pipeline therefore, uses the structured z-only representation, since it preserves physically meaningful deviation patterns while remaining compact enough for prompt construction.

\begin{table}[!t]
\caption{Representative explanation examples from Gemma-3 27B for correct predictions.}
\label{tab:gemma_correct_prediction_explainability}
\begin{center}
\tiny
\renewcommand{\arraystretch}{1.15}
\setlength{\tabcolsep}{2pt}

\begin{tabular}{@{}p{1.35cm}p{2.25cm}p{4.15cm}@{}}
\toprule
\textbf{Case Study} 
& \textbf{Prediction / Fault} 
& \textbf{Model explanation} \\
\midrule

ASM Gasoline Engine 
& \texttt{n\_Engine} / \texttt{n\_Engine}

Engine speed

High confidence
& Gemma-3 27B identifies \texttt{n\_Engine} as the root-cause location. The explanation refers to direct KG evidence from \texttt{n\_Engine[rpm]}, which has the strongest diagnostic score. The calibrated ranking places \texttt{n\_Engine} ahead of Throttle and AccPedal, supporting the diagnosis. \\

\midrule

EV 
& \texttt{BAT\_HV} / \texttt{BAT\_HV}

High-voltage battery

High confidence
& Gemma-3 27B identifies \texttt{BAT\_HV} as the root-cause location. The explanation refers to direct battery-related evidence from \texttt{V\_Bat\_HV[V]}, \texttt{V\_Bat\_HV\_Plant[V]}, \texttt{SOC\_Bat\_HV[\%]}, and \texttt{SOC\_HV[\%]}. These signals are more diagnostic than propagated effects from \texttt{Omega\_EM\_Rear} or Steering. \\

\bottomrule
\end{tabular}
\end{center}
\end{table}

\begin{table}[!t]
\caption{Top diagnostic signal deviations for the selected correct Gemma predictions. Scores are computed from deviations against the time-aligned healthy reference trajectory.}
\label{tab:gemma_explainability_top_signals}
\begin{center}
\tiny
\renewcommand{\arraystretch}{1.12}
\setlength{\tabcolsep}{1.5pt}

\resizebox{\columnwidth}{!}{%
\begin{tabular}{@{}p{2.2cm}p{1.6cm}p{3.3cm}ccccc@{}}
\toprule
\textbf{Case Study} 
& \textbf{Window ID} 
& \textbf{Signal} 
& \textbf{Score} 
& \makecell{\textbf{Raw Mean}\\\textbf{Diff.}} 
& \makecell{\textbf{Mean}\\\textbf{Shift z}} 
& \makecell{\textbf{Max}\\\textbf{Dev. z}} 
& \makecell{\textbf{First}\\\textbf{Abn. Time}} \\
\midrule

ASM Gasoline Engine 
& \texttt{ASM009\_w001} 
& \texttt{n\_Engine[rpm]} 
& 20.000 
& 6971.069 
& 21.782 
& 24.838 
& 20.00s \\

ASM Gasoline Engine 
& \texttt{ASM009\_w001} 
& \texttt{p\_Out\_EGR[Pa]} 
& 4.547 
& -84559.420 
& 4.298 
& 4.921 
& 20.23s \\

ASM Gasoline Engine 
& \texttt{ASM009\_w001} 
& \texttt{p\_Rail[bar]} 
& 1.952 
& -0.057 
& 0.008 
& 4.868 
& 22.44s \\

ASM Gasoline Engine 
& \texttt{ASM009\_w001} 
& \texttt{Pos\_Throttle[\%]} 
& 1.826 
& -56.950 
& 1.402 
& 2.463 
& None \\

ASM Gasoline Engine 
& \texttt{ASM009\_w001} 
& \texttt{Pos\_BrakePedal[\%]} 
& 1.684 
& -1.226 
& 0.368 
& 3.656 
& 22.15s \\

\midrule

EV 
& \texttt{EV001\_w000} 
& \texttt{T\_Bat\_HV[degC]} 
& 15.870 
& 3.237 
& 15.794 
& 15.985 
& 15.00s \\

EV 
& \texttt{EV001\_w000} 
& \texttt{V\_Bat\_HV\_Plant[V]} 
& 5.204 
& -23.049 
& 2.350 
& 9.486 
& 15.00s \\

EV 
& \texttt{EV001\_w000} 
& \texttt{V\_Bat\_HV[V]} 
& 5.204 
& -23.049 
& 2.350 
& 9.486 
& 15.00s \\

EV 
& \texttt{EV001\_w000} 
& \texttt{V\_DCLink[V]} 
& 5.202 
& -23.063 
& 2.350 
& 9.481 
& 15.00s \\

EV 
& \texttt{EV001\_w000} 
& \texttt{I\_Bat\_HV[A]} 
& 3.651 
& -50.653 
& 0.684 
& 8.101 
& 15.00s \\

\bottomrule
\end{tabular}%
}

\vspace{1mm}
\parbox{\columnwidth}{\scriptsize Abn.: abnormal; Dev.: deviation.}
\end{center}
\end{table}

The explanation examples in Table~\ref{tab:gemma_correct_prediction_explainability} show that the LLM decisions are consistent with the KG-guided evidence. In the ASM case, Gemma-3 27B correctly predicts \texttt{n\_Engine} and refers to the direct signal \texttt{n\_Engine[rpm]}, which also has the strongest deviation score in Table~\ref{tab:gemma_explainability_top_signals}. In the EV case, the model correctly identifies \texttt{BAT\_HV} and supports the decision using battery-related signals such as \texttt{V\_Bat\_HV[V]} and \texttt{V\_Bat\_HV\_Plant[V]}. These examples illustrate the benefit of combining numerical evidence with KG relations: the model can distinguish likely source signals from downstream propagation effects, instead of treating all abnormal signals as equivalent.

\subsection{Computational Complexity and Accuracy--Latency Trade-off}

\begin{figure}[htbp]
\centering
\includegraphics[width=\columnwidth]{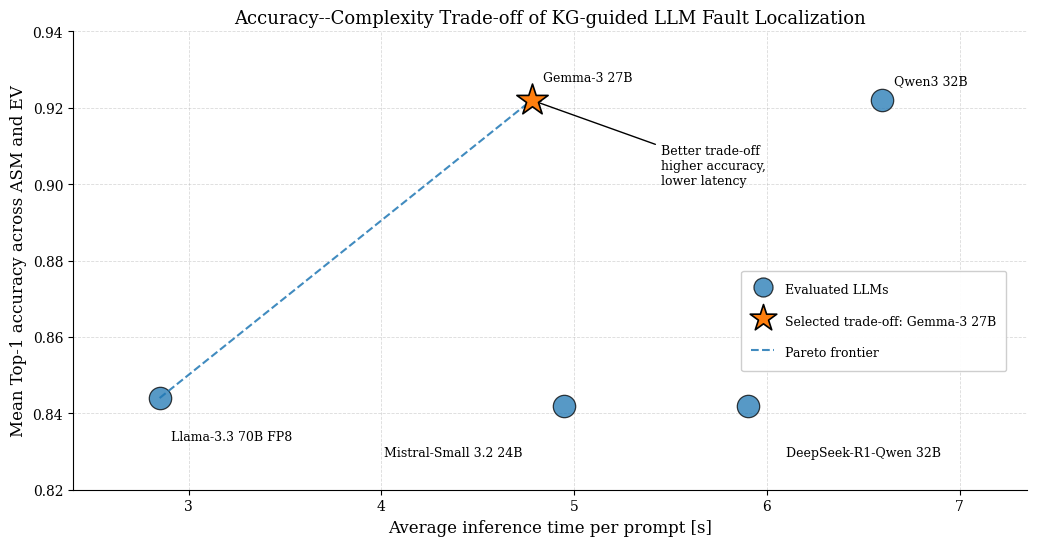}
\caption{Accuracy--complexity trade-off of the evaluated LLMs for KG-guided fault localization. Gemma-3 27B achieves the same highest mean Top-1 accuracy as Qwen3 32B while requiring lower average inference time, making it the preferred accuracy--latency trade-off.}
\label{fig:accuracy_complexity_tradeoff}
\end{figure}

The practical usefulness of the proposed KG-guided RAG pipeline depends on both localization performance and computational cost. The LLM experiments were executed using a vLLM serving setup on an NVIDIA RTX PRO 6000 Blackwell GPU with 97{,}887~MiB of available GPU memory, CUDA~13.2, and driver version~595.71.05. Table~\ref{tab:llm_resource_complexity} summarizes the resource measurements of the evaluated models. Inference time and token counts are averaged over the evaluation prompts, while peak memory denotes the total GPU memory reserved during inference on the evaluation hardware.

\begin{table}[htbp]
\caption{Resource and complexity comparison of the evaluated LLMs using the KG-guided few-shot fault-localization prompt. Values are averaged over the evaluation prompts.}
\label{tab:llm_resource_complexity}
\begin{center}
\tiny
\renewcommand{\arraystretch}{1.12}
\setlength{\tabcolsep}{1.3pt}

\resizebox{\columnwidth}{!}{%
\begin{tabular}{@{}p{3.1cm}ccccccc@{}}
\toprule
\textbf{Model} 
& \makecell{\textbf{Inf.}\\\textbf{time [s]}} 
& \textbf{Thr.} 
& \makecell{\textbf{Prompt}\\\textbf{tokens}} 
& \makecell{\textbf{Output}\\\textbf{tokens}} 
& \makecell{\textbf{Load}\\\textbf{time [s]}} 
& \makecell{\textbf{Mem.}\\\textbf{load [GB]}} 
& \makecell{\textbf{Peak}\\\textbf{mem. [GB]}} \\
\midrule

Gemma 3 27B 
& 4.773 
& 0.210 
& 1502.644 
& 101.056 
& 206.055 
& 86.013 
& 86.013 \\

Qwen3 32B 
& 6.614 
& 0.152 
& 1502.644 
& 110.087 
& 174.051 
& 86.241 
& 86.243 \\

DeepSeek R1 Distill Qwen 32B 
& 5.879 
& 0.171 
& 1502.644 
& 92.515 
& 82.025 
& 86.271 
& 86.271 \\

Mistral Small 3.2 24B 
& 4.940 
& 0.203 
& 1502.644 
& 112.817 
& 108.034 
& 86.353 
& 86.353 \\

Llama 3.3 70B FP8 
& \textbf{2.856} 
& \textbf{0.351} 
& 1502.644 
& \textbf{68.039} 
& 84.023 
& 86.333 
& 86.333 \\

\bottomrule
\end{tabular}%
}

\vspace{1mm}
\parbox{\columnwidth}{\scriptsize Thr.: throughput. Prompt-token counts are computed with each model's native tokenizer. Peak memory denotes the total GPU memory reserved during inference, not the minimal model footprint.}
\end{center}
\end{table}

The results show a clear accuracy-latency trade-off. Llama-3.3 70B FP8 is the fastest model, requiring 2.856~s per prompt and reaching the highest throughput of 0.351 prompts/s. However, its localization performance is lower than the best-performing models, especially on the ASM gasoline engine case. In contrast, Qwen3 32B and Gemma-3 27B achieve the strongest predictive performance across both case studies. Both models obtain the best overall Top-1 accuracy, Top-3 accuracy, MRR, F1-score, and MCC values. Their runtime differs, however: Qwen3 32B requires 6.614~s per prompt, whereas Gemma-3 27B requires 4.773~s per prompt. Thus, Gemma reaches the same highest localization performance while reducing the average inference time by approximately 28\%. Fig.~\ref{fig:accuracy_complexity_tradeoff} visualizes this trade-off between mean Top-1 accuracy and inference time. Gemma-3 27B lies on the Pareto frontier because no other evaluated model is both more accurate and faster. Although Llama-3.3 70B FP8 is computationally efficient, its lower accuracy makes it less suitable as the final diagnostic model. Qwen3 32B provides the same accuracy as Gemma, but at a higher inference cost. Since model loading is a one-time cost during a serving session and GPU memory usage is similar across models at approximately 86~GB, the average inference time per prompt is the most relevant runtime criterion for the offline HiL analysis workflow. Gemma-3 27B provides the most favorable accuracy-latency trade-off and is selected as the preferred model for the proposed KG-guided RAG-based fault-localization pipeline.

\section{Conclusion}
\label{sec:conclusion}

This paper presented a KG-guided retrieval-augmented LLM framework for root cause analysis and fault localization in automotive HiL validation data. The proposed method transforms raw multivariate recordings into healthy-reference diagnostic evidence, enriches this evidence with sensor-to-location and propagation knowledge, retrieves similar historical cases, and uses an LLM as a reasoning layer to generate ranked fault-location predictions with explanatory RCA output.

The framework was evaluated on two HiL case studies: an ASM gasoline engine and an electric vehicle system. The results show that structured diagnostic evidence, KG-guided candidate ranking, and retrieval-augmented prompting can support reliable and explainable fault localization across different vehicle architectures. In particular, the strongest models achieved high Top-1 localization accuracy, while recording-level aggregation improved the stability of the final diagnosis by reducing the influence of locally ambiguous windows.

The current study has some limitations. The evaluation dataset is still limited in size and mainly focuses on gain faults. In addition, the present experiments address single-fault localization, while concurrent faults may create overlapping direct and propagated effects that are more difficult to separate. The quality of the RCA explanation also depends on the completeness of the knowledge graph and the available historical case library.

Finally, future work will extend the framework to concurrent and interacting fault scenarios, larger HiL datasets, and additional fault types. Another direction is to integrate richer spatial or causal representations, such as dynamic knowledge graphs or graph neural networks, to model fault propagation more explicitly.

\bibliographystyle{IEEEtran}
\bibliography{Reference}

\end{document}